# Communicating astronomy with the public: perspectives of an international community of practice

Sara Anjos, Pedro Russo and Anabela Carvalho

**Abstract**  Communities of practice in science communication can make important contributions to public engagement with science but are under-researched. In this article, we look at the perspectives of a community of practice in astronomy communication regarding (relations with) their public(s). Most participants in this study consider that public(s) have several deficits and vulnerabilities. Moreover, practitioners have little to no contact with (and therefore make no use of) academic research on science communication. We argue that collaboration between science communication researchers and practitioners could benefit the science-public relationship and that communities of practice may be critical to that purpose.

**Context**  Science and Technology (S&T) are at the core of contemporary culture, welfare and democracy. S&T are expected to increase economic growth and improve people's lives in many ways, and shape citizens' identities and thinking regarding societal issues such as climate change, energy, food security and health [National Academies of Sciences, Engineering, and Medicine et al., 2016; Davies and Horst, 2016].

Such expectations put the relations between the scientific community and (the rest of) society under the spotlight and increase the importance of individuals and organizations working as facilitators (or translators) of science. This is not an easy task. The complex environment of modern societies — where different specializations, cultures, languages and modes of meaning-making coexist — adds to the challenges of developing science communication practices that may serve multiple publics and their diverse needs.

Science communicators "move knowledge around" [Meyer, 2010, p. 118], and also create new forms of knowledge [Meyer, 2010]. As key actors in the establishment of connections "between science and society at large, making elements of the science domain approachable, understandable and eventually appealing" [Bucchi and Trench, 2014, p. 2], they are essential to implement activities towards public engagement with science at various levels and may even operationalize science policies driven by political agendas [Weingart and Joubert, 2019].

In order to foster connections and mutual understanding between research communities and other sectors of society, science communicators aim to develop a "common language" in their practice [Meyer, 2010; Stocklmayer, Gore and Bryant, 2001]. Practitioners of science communication may simultaneously be members of multiple professional groups and take up different roles, such as scientists, informal science educators, press officers, science journalists, bloggers, and others. It is often the case that science communicators organize themselves into networks of knowledge and experience sharing, i.e. into communities of practice. These are composed of (groups of) individuals that gather around a common goal: the wish to improve communication practices brings the community together. Practitioners act within their community of practice according to their institutional role and develop various forms of interaction thereby aligning perspectives and facilitating transactions among multiple stakeholders [Kuhn, 2002; Meyer, 2010; Wenger and Wenger-Trayner, 2015].

Several studies have focused on scientists' participation in science communication [e.g. Bauer and Jensen, 2011; Besley and Nisbet, 2013; Entradas and Bauer, 2019; Wellcome Trust, 2001]; however, research that focuses primarily on practitioners' perceptions, and especially on the roles of this community of practice, is scarce. Hence, the contribution of these specific social groups and structures to the science-society relationship has been overlooked.

Aside from this research gap, there seems to be a certain distance between the practice of science communication and the corresponding academic research [Miller, 2008; Riesch, Potter and Davies, 2016]. On the one hand, the normative ways in which public understanding of science (PUS) research often addresses science communication matters neglect the body of reflections and knowledge that practitioners develop in their practice; on the other hand, that research is seen as of little use for practitioners[1] [Riesch, Potter and Davies, 2016]. Although science communication practitioners and academics are distinct communities, narrowing the relationship between research and practice in science communication is a continuously important goal [Seethaler et al., 2019]. In fact, there is evidence that when this kind of collaboration between research and practice on science communication occurs, multiple benefits are achieved, as reported by Riedlinger et al. [2019] with regard to storytelling for engaging with the public. Such collaboration can "also produce generalisable findings and contribute to theory building in the science communication field" [p. 10]. Therefore, furthering dialogues between science communication research and practice [Jensen and Gerber, 2020], bringing practitioners' experiences to peer-reviewed literature,[2] and linking empirical scholarship to practice are expected to promote mutual gains, and unleash diverse opportunities for improvement concerning a variety of topics and methods for communicating science. This kind of cooperation may be particularly encouraged within communities of practice, which typically gather multiple groups of people around common interests and experiment different approaches. Moreover, communities of practice may use and spread academic research contributing to bridge the scholarship-practice divide.

---

[1]See the report on the Rockefeller Science Communication Conference, at The Bellagio Center, Italy, 6–10 November 2017, Topic 3: Science Communication Practitioner "The world of the science communication practitioner" in https://www.scicom-bellagio.com/2017/12/20/bellagio-the-world-of-the-science-communication-practitioner/ (visited on 18 April 2019).

[2]For instance, Canfield et al. [2020] have noticed that several practitioners are experimenting with methodologies for inclusiveness in science communication that have not yet been addressed in academic literature.

In line with the points that we are making, the systematic review of science communication research conducted by Gerber et al. [2020] identified the lack of knowledge transfer between theory and practice as a challenge for the area. They also found a gap concerning research on science communication practitioners themselves. This article offers contributions to start filling these gaps.

**Objectives**

This study analyses the perceptions and practices of a specific community in science communication, namely the astronomy communication community, which is briefly characterized in the next paragraph.

The International Astronomical Union (IAU) was founded in 1919 "to promote and safeguard the science of astronomy in all its aspects, including research, communication, education and development, through international cooperation".[3] To pursue its objectives, the IAU created several Divisions (D), Commissions (C) and Working Groups (WG). Among them, we find the Commission C2 — Communicating Astronomy with the Public,[4] which brings together a worldwide group of individuals working on public communication of astronomy. The Commission functions as a hub for members with the common interest of communicating astronomy to the public, and provides opportunities for the community to share resources and best practices via several platforms, such as the practitioner journal *Communicating Astronomy with the Public (CAP)*[5] and the biannual CAP conference. Described as a "think/do tank that convenes the astronomy communication community, and seeds initiatives to explore new ways to communicate astronomy with the public", the conference is a privileged encounter of the astronomy communicators' community of practice aiming to "endorse standards, best practices and requirements for public communication".[6] The 2018 CAP conference took place in Fukuoka, Japan, and had 446 participants from 53 countries. The gathering provided an opportunity to study practitioners' perceptions about public engagement with astronomy and interconnections of science communication research and practice.

We explore the following questions:

1. How do astronomy communication practitioners conceive their public(s) and which impacts do they expect to have?

---

[3]Retrieved from https://www.iau.org/administration/about/ (visited on 20 September 2019).
[4]Until 2015, the commission was called Commission 55. For more details on the commission's description and objectives: https://www.iau.org/science/scientific_bodies/commissions/C2/.
[5]Retrieved from https://www.capjournal.org/ (visited on 20 September 2019).
[6]Retrieved from https://www.communicatingastronomy.org/about/ (visited on 1 March 2018).

2. What kind of public engagement do practitioners seek and how is that embedded in their interventions?

3. How do they appropriate science communication research in their practices?

This work results from the authors' multiple lenses: on the one hand, as elements of the community of practice in astronomy communication (the two first authors) and, on the other hand, as researchers in the field of science communication. Supported by diverse methods of data collection, we conducted an exploratory analysis of the community's perspectives, as detailed below.

**Methodology**

A key starting point for this study is the assumption that social subjects' interpretations and the meanings that they ascribe to reality are essential to inform social research.

In looking at community practices, we used a participant observation approach, which focus on meanings as seen from the standpoint of insiders [Flick, 2004; Hammersley, 2015]. In order to acquire detailed insights into participants' experiences and perspectives, we conducted a set of semi-structured interviews [Jensen and Laurie, 2016]. In addition, we used other data sources, namely document analysis (of certain aspects of the conference proceedings book) and a field diary. These methods allowed us to complement the information offered in the interviews, showing us how practices really work [Flick, 2004].

The question of representativeness was present in our multiple discussions of the study: can we present findings in terms of the community or can we only speak of the perceptions of some of its members? Although we acknowledge that communities of practice are varied and formed by individuals with different profiles, we realized that certain ideas and perspectives kept appearing recurrently in the interviews. As we neared 'meaning saturation', we considered that more interviews would not have changed our findings substantively. Observation of conference presentations, informal conversations and other forms of interaction of community members further contributed to identifying shared views. What we offer below is thus the result of multiple methods for collecting and analysing data that led to an interpretative and critical analysis of practices and meanings. We focus on what appears to be most common within the astronomy communication community, although care should be exercised in appraising this analysis to avoid over-generalizations. Science communication is always situated, and is motivated by diverse reasons and purposes [Aikenhead, 2001; Canfield et al., 2020; Entradas, 2016], which requires looking at the specific contexts where it happens.

We selected 16 interviewees ($N = 16$; 5 female, 11 male) during the conference, including practitioners from different countries, at different career stages and, working with different audiences (see Table 1). Most interviewees ($N = 12$) held a university degree in the field of Physics or Astronomy, and four reported having some advanced training in informal science education or science communication. For the selection of respondents, 10 community active members were initially identified. To this end, we took the following factors into account: participation in previous editions of the CAP; membership of the C2 committee and other IAU groups of interest to the community (such as the WG Astronomy for Equity and Inclusion, the WG Astronomy Educational Resources — AstroEDU, the Office of Astronomy for Outreach, and the Office of Astronomy for Development); and/or involvement as members of the 2018 conference organizing committee. The remaining interviewees were suggested as interviews progressed and were selected considering their age, career stage, function, geographical location, and the type of audience to which they address their practices. A strong presence of interviewees from Asia reflected the distribution of conference participants: 277 came from Asia, of which 200 came from Japan, the conference venue. The interviews were anonymous in order to reduce possible restrictions on responses, associated with constraints related to name, institution, function or position as astronomy communicators.

Table 1. Respondents profiles and audiences.

| | | E1 | E2 | E3 | E4 | E5 | E6 | E7 | E8 | E9 | E10 | E11 | E12 | E13 | E14 | E15 | E16 | Total |
|---|---|---|---|---|---|---|---|---|---|---|---|---|---|---|---|---|---|---|
| *Gender* | Masculine | × | | | | × | | × | × | | × | × | × | × | × | × | × | 11 |
| | Feminine | | × | × | × | | × | | | × | | | | | | | | 5 |
| *Age (years)* | 25–34 | | | | × | × | | | | | × | | | | | | × | 4 |
| | 35–44 | × | | × | | | × | × | | × | | | × | | × | | | 7 |
| | 45–54 | | × | | | | | | × | | | | | × | | × | | 4 |
| | > 55 | | | | | | | | | | | | × | | | | | 1 |
| *Geographic location* | Africa | × | | | | | | | | | × | | | | | | | 2 |
| | Asia | | | × | × | × | × | | | | | | × | | | × | × | 7 |
| | Europe | | × | | | | | × | | | | | | × | | | | 3 |
| | North America | | | | | | | | | × | | | | | | | | 1 |
| | Oceania | | | | | | | | | | | | | | × | | | 1 |
| | South America | | | | | | | | × | | | | | | × | | | 2 |
| *Main audience* | Primary and secondary school students | × | × | | × | × | | | | | | × | × | × | | × | × | 9 |
| | University and college students | × | | | × | | | | | | | | | | × | × | | 4 |
| | General public | × | × | × | | × | | × | × | | | × | × | × | | × | | 10 |
| | Media | × | | × | | | × | | | | | | | | × | | | 4 |
| | Internet users | | | × | | | | × | | × | | | | | | | | 3 |
| | School teachers | | | | × | | | | | | | × | × | × | | × | | 5 |
| | Outreach and education actors, astronomers | | | | | | × | | | | × | | | | | | | 2 |
| | Policymakers | | | | | | | | × | | × | | | × | | | | 3 |

Despite the practitioners' identification of several publics, it was possible to realize that more than half developed their main activity, directly or indirectly, in formal and informal educational settings (with pre-University students, teachers and educators). Four interviewees spoke of the media as a significant audience. It should also be noted that two individuals worked in the tourism sector (where there has been a growth in the offer of experiences of astronomical observation).

Table 2. Goals, questions, and themes regarding the interviews.

| Research goals | Interview questions | Key themes identified |
|---|---|---|
| To understand practitioners' perceptions of the public | Q1. What is/are your main audience(s) and where do you normally communicate science/astronomy?<br>Q2. How do you see the science-public relationship? Do you think the astronomy-public relationship is different in any way? What do you think the public wants/needs/values? | The public |
| To analyse practitioners' perceptions of their role in public engagement | Q3. What do you think your main roles are (when communicating science) in relation to public understanding and engagement with astronomy? Why? | Public engagement with science: major concerns |
| To reflect on practices of astronomy communication | Q4. Please describe a specific astronomy communication activity that you have been involved with recently and the different phases of the activity — design, implementation, evaluation | Practices for engagement |

Table 2 presents our research goals regarding the interviews in connection with the interview script. The script was adjusted while conducting the interviews with the various participants to acknowledge their spontaneous references to some themes and to enhance the depth of responses (for instance, by asking for specific examples or drawing on something that was said and asking for elaboration). Interviews ranged from 15 to 50 minutes and were audio-recorded after verbal permission. Further to transcription, we analysed the data assisted by RQDA software [Huang, 2014]. We then constructed 33 codes, which after additional investigation led to three main themes: the public; the community's concerns regarding public engagement; and practices for engagement (see Table 2).

We also considered the field diary notes collected throughout the conference (concerning observations at workshops, and parallel and poster sessions) with the aim of searching for common goals, methodologies, and narratives among the community.

Finally, we analysed the conference proceedings book and looked for references to academic publications on science communication with the goal of understanding how that knowledge may be considered and appropriated by astronomy communicators.

**Findings**

In this section, we present findings associated with the main themes found in the interviews, complemented with insights from observation of practices and with analysis of the CAP proceedings book. Regardless of the diversity of respondents and conference participants (the community is geographically spread, targets different audiences and communicates on behalf of different stakeholders), we found several common views and practices across the community.

*Views on the public*

Understandings of the public(s) of science communication are vitally important to practice. Below we discuss several traits that emerged from our interviewees.

**The public is emotionally connected to astronomy.** Practitioners consider that the public values the science of astronomy in seeking answers to their questions and appreciate those who can elucidate them. Astronomy, they claim, offers "*magestical*" pictures (namely from NASA and ESO) and addresses the big questions ("*Where do we come from? Where are we going? What will happen to our home, the Earth? What is our place in the universe? Are we alone?*" [E6]). Besides, there is some "*magic and romance*" in the stars and planets ("*we wish upon a star*" [E9]), contributing for people to have an affective liaison to the subject. The public is, therefore, portrayed as emotionally connected to astronomy, which can be a powerful hook to trigger interest for science in general, especially STEM (Science, Technology, Engineering and Mathematics) subject areas.

**The public lacks knowledge and has several misconceptions and stereotypes about science and scientists.** Most interviewees ($N = 13$) mentioned that the public lacks knowledge of science and the scientific process and holds several misconceptions, which are particularly damaging in face of various forms of pseudoscience. For astronomy communicators, the confusion between astronomy and astrology is a much-referenced example. In this respect, the lack of scientific literacy to distinguish science from pseudoscience is seen as critical [Allum, 2011; Stocklmayer, Gore and Bryant, 2001].

According to interviewees ($N = 11$), the public associates several stereotypes with science and scientists: "*science is difficult*", "*[it is] only for special people*" [E16]. Some referred to concrete actions to combat stereotypes, such as gender-related ones: "*I always try to use examples of the female astronomers in my talks*" [E13]. The mentioned barriers for access to science also included racial, ethnic, disability or socio-economic conditions. In contrast, interviewees emphasized that astronomy is inclusive and suited to all, as it addresses fundamental questions that concern the whole of humanity: "*Astronomy is a science that can be reached by everyone through inclusion, through diversity, through equity policies, through empathy*" [E6]. In fact, the idea that astronomy brings improvements to society ("*the importance of astronomy for the betterment of society*" [E6]) is a recurrent assumption in the speech of community members, who seem personally committed to its promotion.

**The public may influence science policies.** Several interviewees ($N = 7$) recognized that public support is essential to influence science policies, especially science funding. Science policies and the role of governments and scientific organizations were highlighted as crucial aspects of the science-society relationship. Policies were prized either by potentially promoting equal access amongst several groups (considering gender, ethnicity, race, disabilities), by promoting proximity and free access to science, or by levering the economy through investments on science infrastructures.

Although astronomy communicators saw the public as very important in relation to science policies, we found no considerations in the interviews that would lead us to think that, in their opinion, the public should be a participant in the governance of science. This was not even the case in the speech of practitioners based in Europe where there is a substantial investment in relevant programs, such as the European Union's framework programme Horizon 2020 Science With and For Society (Swafs),[7] which is based on democratic ideals. Communication practitioners tend to assume that people have the right to know where their money and taxes are being spent, and to know what are the products and applications for society of the investment in science, but this is how far views on science governance and democratization go. Public engagement understood as participation in policy seems to remain confined to abstract discourses in politics and academia, with the astronomy communication community paying little attention — and assigning only a limited role — to the public in policymaking [Entradas, 2016].

*Practices for public engagement*

Astronomy communicators see themselves as facilitators between science and the public, with the tools and methodologies to bring science to the public being planned in a way that matches people's interests and expectations. The community appears to respond to the public's perceived emotional connection to astronomy by using emotional content (namely scientific storytelling), and by promoting empathy towards science and scientists (scientists are "*just humans*" [E7]). They also emphasized the relevance of astronomy in people's daily life and in society, and how it may help solve societal issues.

In order to change the public's attitude towards science, communicators point them to "*reliable sources*" about the science of astronomy and the processes involved in knowledge-building in order to build a "*scientific mind*" and "*critical thinking*" [E3]. Communicators described a diverse range of actions, both in form and content, but the predominant model of communication with the public rests on the idea of "transmission" of information about science and scientists ("*this is the message I have to transmit*" [E7]; "*that will make sure that the message gets across*" [E13]). This prevailing traditional view, based on the transmission model, according to the theory of communication [e.g. McQuail, 1998], indicates the continuing dominance of a one-way communication model for enhancement of scientific literacy [Bauer, 2009; Pardo and Calvo, 2004].

Nevertheless, we found clues in practitioners' speech to spaces for *quasi*-dialogue and discussion about science subjects. The perception that the public enjoys to converse about and discuss science, and knowing the daily life of a scientist and science institution, shows that the public seeks a relationship of trust, closeness and transparency ("*the public is very interested in what we do*" [E7]). Closer proximity to scientists and scientific institutions are associated with a positive attitude by the public. Although not clear about the terms in which dialogue may occur, interviewees seemed to suggest that it is mostly related to knowledge of facts and processes of science, and to returns on investment in science and finance:

> *"everything nowadays has to have an economic value, [...] every time you talk about astronomy then you have one person in the group asking: '— yeah but what is the output, what can we sell out of astronomy?'"* [E7].

---

[7]For more detail, please visit https://ec.europa.eu/research/swafs/index.cfm (visited on 17 February 2019).

Although interviewees employed the term "*engagement*" several times, it seemed to refer to people's involvement with science in a cognitive, behavioural or emotional sense (aspects that tend to be foregrounded in education research) and not so much in the sense of participation and governance of science (more commonly discussed in the field of science communication research) [e.g. Lewenstein, 2015].

The astronomy communication community seems to conceive this cognitive and emotional engagement as the first step for science literacy. In their responses to our questions, practitioners offered clues to multiple goals for engaging the public with astronomy, namely:

- to increase interest in and awareness of science;
- to talk about the process of scientific knowledge construction;
- to make people aware of the practices and constraints of science and scientists;
- to demonstrate the applications of science in the practical life of citizens (Global Positioning System (GPS), charge-coupled device (CCD), etc.);
- to make connections between astronomy and other areas of knowledge and entertainment (STEM, History, Music, Art);
- to talk about astronomy's contributions to solve societal problems (global warming, waste, energy, health);
- to help develop critical thinking for assessing the trustworthiness of scientific claims (mainly in the media).

All these aspects support the construction of critical science literacy [Carvalho, 2004; Priest, 2013], which accounts for broad questions related to the production of scientific research and its links to several contexts of people's lives.

Despite interviewees referring to motives for concern related to the education and media arenas, we found that the astronomy communication community aspires to have education and media agents as allies in promoting interest in science and helping address several of their priorities.

*Major concerns on engagement*

Interviewees pointed to various types of concerns relevant to their public engagement practices. They mainly concern science at school, STEM careers, the gender gap, and the media.

Interviewees corroborated the idea that students' dislike for science at school is one of the reasons for low rates in academic and professional careers linked to science [DeWitt, Osborne et al., 2013] and aimed for attitude change. They consider that they can play a part by providing resources and fun activities:

> "*at school, I think kids tend to say they don't like science or they hate math*"; "*once they get out of the school and come to our science centre and they have lots of fun things, they usually enjoy science*" [E14].

Another concern is the gender gap on STEM subjects, and how astronomy may contribute positively [e.g. Sjøberg and Schreiner, 2010]. Some studies [e.g. DeWitt and Bultitude, 2020; Lane, Goh and Driver-Linn, 2012] have pointed out that the link between science, power and male roles can be an obstacle to some disadvantaged groups, including women, indigenous and African-descent communities, which some of the practitioners interviewed in our research also mentioned. This raises questions about the ways in which science literacy has been problematized to contribute to attitude change among these specific social groups and how communities of practice are addressing this concern. Although the astronomy communication community seemed open and alert to this kind of issues, members showed some uncertainty on how to direct their practices. One interviewee suggested looking into social sciences research on matters such as gender balance and inclusion: "[we have] *a specific section on diversity and inclusion in a more formal research on social sciences*" [E6].

The media also seem to represent a concern to this community. Despite being seen as essential to disseminate scientific information and to raise public interest in science, interviewees consider that the media reinforce existing stereotypes and misconceptions related to science and scientists, and that media professionals do not always bother to convey the correct information by checking sources:

> "*some people view, — and sometimes the media views — scientists like geniuses*" [E14];
>
> "*usually I write [a news story] myself, and I give it for the media, [or else] [. . . ] there is a wrong interpretation and so I don't want to have that*" [E1].

They attribute an important role to the media in shaping opinions about science and scientists ("*the media has a very important role because they are also the ones who define people right now*" [E3]). However, interviewees affirm that there is a barrier between the media and the scientific community, a mutual misunderstanding and mistrust, which some academic studies also demonstrate [e.g. Besley and Nisbet, 2013; Peters et al., 2008]. Suspicion is stronger concerning new media in online platforms than the traditional press, which frequently has (or used to have) specialized science journalists:

> "*we have the reporter or journalist for print media, they have their rules, and they stick to the rules up until now. I know them, I can say that because I know them, I work with them, the problem is in online media*" [E3].

We noticed an attempt to work with the media, so that media professionals may have a better understanding of science, scientific facts, science processes and scientists. However, once again, the vast majority of interviewees refer to the media and the communication process in a simplistic, linear, sender/receiver-type model ("*the information that they [the media] are going to transmit*" [E1]). We found no reference to how people interpret and infer meanings (in increasingly multimodal forms of communication) and understand scientific discourses in the media in real-life, contextual settings [e.g. Basu and Barton, 2010; Kress, 2003].

*Research-practice divide*

The community regularly shares experiences as a means for identifying, monitoring and enhancing good practices, in other words, as a kind of self-regulation and peer-review practices. Our observation of the conference, where most presentations focused on members' "do's and do not's", reinforced this idea. Analysis of the proceedings book references section showed many links to organizations and project websites but rather few to academic publications (see Table 3). This confirms that academic research has little to no weight on the community's practice and should prompt further study to explain the motives behind this fact.

Personal experiences and "lessons learned" steer a large part of the community's activities, with little objective evidence of their efficacy or impact. Evaluation seems to be left to a secondary plane, even though practitioners recognize its importance to improve practice. Science communication scholars have put forth several proposals to do so in a systematic manner [Falk and Needham, 2011; Fischhoff, 2019; Jensen and Gerber, 2020], which could create collaboration opportunities between research and practice in science communication. Projects involving storytelling, where this kind of cooperation has shown several advantages to research and practice likewise, may also offer interesting opportunities [Riedlinger et al., 2019]. Yet another example is research on emotions and emotional appeals in the context of science communication, which suggests that they can be an instrument to promote action and a significant predictor of perceived trustworthiness [Reif et al., 2020; Taddicken and Reif, 2020]. This may be useful to confirm if interest (e.g. triggering interest in STEM) and trust are being enhanced by this community. These kinds of collaborations may be useful to guide and adjust action plans, and likely result in better ways of relating to the public(s).

Table 3. Analysis of the references section of the CAP 2018 proceedings book.

| Number of presentations (total) | Number of presentations | | |
| --- | --- | --- | --- |
| | With no references section | With references section, but no references to academic research on science communication or science education* | With at least one reference to academic research on science communication or science education |
| 341 | 125 | 160 | 56 |

\* The references section included links to organizational websites, project websites and personal ones. We also found references to symposia and proceedings of conferences of the community, which were not found in a Google Scholar search, hence not easy to access by those outside the community.

Looking more deeply at the issues of concern to the community, which are frequently discussed in communication research, may be beneficial to the practice of astronomy communication. Reflecting on communication models, symbols and logics used by the media would probably be helpful to promote change in public attitudes regarding misconceptions and stereotypes identified by many of the interviewees. This becomes even more relevant with regard to younger audiences, as they make a constant use of online communication platforms and are, therefore, potentially more exposed to information and opinions about science that circulate in those media, which are more difficult to follow or control.

Several scholars have analysed the potential to promote public debate and participation in scientific processes through social media platforms [Bik and Goldstein, 2013; Brossard and Schefeule, 2013]. This is also a way for scientists to make their work and opinions known [Bik and Goldstein, 2013], and to become influencers and opinion makers on science subjects. Although the media were constantly referred to, we did not find clues that allowed us to think that, in general, this community of practice is using media to promote discussion and participation in science matters. In fact, the work of Entradas and Bauer [2019] showed that astronomers make little use of social media, and many of the CAP network members identify themselves as astronomers. Mostly, the media's role was confined to publicizing scientific events (such as talks or astronomical observations, for example), disseminating scientific content and enhancing organizational communication (of science institutions). Many studies focus on scientists' skills to use and comprehend media [e.g. Peters, 2013; Peters et al., 2008; Pinto and Carvalho, 2011] but it is necessary to continue examining these matters in a context where online media have assumed a greater role in the selection of the scientific sources of information that people consume [Brossard and Schefeule, 2013].

The community is skilled in the use of multimedia tools on different platforms and certainly has the potential to use media to foster participation. However, their involvement in dialogical and participatory forms of communication can be much improved.

In our view, citizen science projects have a great potential for public engagement in the sense of democratic participation, as well as for greater interconnection between research and practice in science communication. As put by Lewenstein [Gerber et al., 2020, p. 39]:

> "Citizen science is one of the areas where the boundaries of science communication research and practice are getting more diffuse and dissolving because people in that world are trying to ask questions about motivation, recruitment and outcomes to improve their practice."

However, we found a thin representation of this type of project at the CAP conference with only 5 articles and 8 posters referring to citizen science initiatives in the proceedings book. Some of these make use of social media to encourage people to contribute to the science of astronomy (e.g. by sending photographs or extracting scientific results) and it seems to us that it can be leveraged for a greater diversity of citizen engagement in science.

**Conclusions**

Based on this exploratory study, we can draw several conclusions and reflect on interconnected aspects that are worthy of further research. Firstly, most interviewees in the astronomy communication community tended to homogenize the public and to emphasize deficits and vulnerabilities of several kinds (for instance, the ideas that the public has only a basic knowledge about science, that they have/had little interest in science while at school, and that they are strongly influenced by the media). Most seemed to disregard various factors that may make people's relationship with science vary. The public is hence mostly seen as having a single role, that is, of a recipient of science information (except concerning science policies, which practitioners admit the public may influence). Given the perceived multiple deficits of the public [Bauer, 2009; Trench, 2008; Bauer, Allum and Miller, 2007], concerns with the risks of misunderstandings and misinformation rise, as well as the perceived need to defend science from such risks [Bucchi and Trench, 2014; Dudo and Besley, 2016]. For a generally unknowledgeable, passive and influenceable audience, the prevailing focus is on the need to transmit the "right message of science", which is expected to change the public's attitude towards science and increase interest in science. In addition, showing scientists and science in action to the public appears to be perceived as an important strategy for communication practitioners to re-establish the "truths" of science and foster an emotional connection. By seeing the behind-the-scenes of the scientific enterprise the public would presumably be more interested and supportive.

Therefore, in the interviewees' speech, we identified two types of functions of the community: a) dissemination of scientific knowledge in order to fill the public's knowledge deficit; b) bridging the gap between the public and the scientific community, which would allow for a better understanding of the processes of production of scientific knowledge and an increase of interest in science. There is a recognition from the community that the public's positive attitude towards science enables science to advance, either by promoting research funding or by attracting potential new scientists. Insofar as they perceive the public as influential with regards to science policies, they constitute the members of the public into citizens that function as political agents in democratic societies. The kind of scientific citizenship they expect from their public frames the kind of engagement they promote with science, which is aligned with existing structures of knowledge production and involves only a low level of dialogue.

Although this community of practice recognizes multiple goals for public engagement, many of which are critical to scientific citizenship, it seems that astronomy communicators are still discovering different forms of relating to their audiences. Awareness of, and reflection on, the multiple ways of assigning meanings to science-related issues are ongoing processes and need to be further encouraged. This calls for cultivating linkages between scientific specialties and other subject areas, as well as analysis of social, political and cultural aspects in each context. The transition of science communication practices to "knowledge building" models requires considering the public as an active co-constructor and user of scientific information [Falk and Dierking, 2012; Stocklmayer and Rennie, 2017]. Therefore, the notion that astronomy communication practitioners tend to hold of their audiences seems to limit the relationship they build with the public(s) of science.

Secondly, we observed that as school and media are viewed as sources of some misconceptions and stereotypes, the community tries to liaise with teachers and journalists on a regular basis so that those agents may "transmit" the "message" of science more accurately. Recognizing the critical role of the media in framing science issues and influencing public perceptions and attitudes, the community shows concerns about media distortion of science overall, affecting science credibility. Trust and credibility have for long been a concern for science communication research, especially regarding the public's perceptions of the credibility of sources of science information [e.g. Bubela et al., 2009; Weingart and Guenther, 2016]. However, it seems that the community holds a traditional stereotyped perspective that holds science as distinct and separate from society at large [e.g. Bucchi and Trench, 2014], which limits their comprehension of other stakeholders' roles in science communication.

Public trust in science is a critical question that needs to be further explored, especially considering new ways of information circulation, increasingly via social media and other internet spaces where scientists do not always participate. Scientists and science institutions are using "media logics" [Bauer, 2008] in their practices in communicating science to the "end-user", not always knowing how to deal with the "dynamics and potential risks of such engagement" [Bucchi, 2017, p. 891].

Also, the institutional connections we have found in interviewees' profiles points to distinct motives for communicating science, and thus the need to consider other factors involved in the trustworthiness of science. Likewise, the growing trend for private funding in science and the instrumentalization of science in political and economic discourses are changing the symbolic environment and challenging traditional forms of certifying the reliability of science information, and, more generally, the social authority of science [e.g. Weingart and Joubert, 2019]. The astronomy communication community needs to be prepared to face growing challenges regarding an open and democratic view of science governance, which call for expanded debates on potential implications of research applications, research funding and research agendas. In our view, this community may be instrumental in challenging the public to take up other (deeper, far-ranging) roles in science governance in collaboration with other agents.

Thirdly, we observed a limited influence of science communication research on this community of practice. We found interviewees to have a limited scholarly knowledge of science communication theory, such as communication models and processes, and audience research. The lack of references to science communication research in CAP participants' speech and in the CAP proceedings book, as well as the interviewees' profiles and answers, lead us to conclude that this community does not engage with — or is not aware or knowledgeable of — science communication research. We should ask why that is the case. Topics of interest to the community, such as inclusion and gender balance, are frequently studied in the social sciences [e.g. Mitchell and McKinnon, 2019; Osborne, Simon and Collins, 2003]. As science communication activities grow and become a more professionalized activity, practitioners' shortcomings to address complex processes of science communication vis-à-vis diversified audiences upsurge. As most practitioners have formal training in science, there is a need for additional training on science communication, as well as for widening collaborations and partnerships with other science communication stakeholders, such as journalists and scholars.

Notably, we identified a need for training oriented to dialogue and to capacity-building of astronomy communicators to foster public participation [Trench and Miller, 2012]. Science communication research mostly refers to "public engagement" as a relationship based on transparent and dialogical communication between citizens, scientists and policy-makers [de Oliveira and Carvalho, 2015]. In this relationship, communication practitioners, as mediators, are expected to play a critical role. Needless to say, an effective collaboration between research and practice is expected to benefit the science-society relationship. Practitioners' lack of familiarity with science communication research may also be a responsibility of the research community and closed-access publication policies.[8] It is thus likewise important that the science communication research community develops new approaches to engage more broadly with practitioners to foster a wealthy exchange of experiences and lessons learned from both sides. There are studies that suggest the existence of several barriers between science communication researchers and practitioners, related to two very different cultures, which may create some tension (see, for instance, Gerber et al. [2020]). This article aims to contribute to help lower some of these barriers, by looking at science communication practice with the lenses of science communication researchers and helping enable a dialogue between the two fields, which will hopefully contribute to boost debate in the forums of these communities, such as conferences and workshops.

Future research may clarify some of the questions that this exploratory study enfolds. The present study offers a "snapshot" of the astronomy communication community at a given moment in time. Since communities of practice are by nature organic and dynamic structures, by observing them we do not expect to find stable "truths" and unique ways of acting in the face of increasingly complex and diverse contexts for the science and society interface. Nonetheless, as various trends, shared ways of seeing and common practices were observed in this study, it is worth examining them, as we have attempted to do, in order to support the community in developing processes of reflexive science communication practice.

**Acknowledgments** Sara Anjos held a Ph.D. grant (SFRH/BD/123276/2016) co-financed by FCT/FSE/MCTES through national funds.

**Note added.**  A preliminary version of this study was presented at the 'VII Jornadas Doutorais em Comunicação & Estudos Culturais' held in Braga, Portugal, October 14–15, 2019, and is included in its proceedings: Anjos, S., Russo, P. and Carvalho, A. (2020). Comunicar Astronomia: representações do público e implicações para a práxis. In Z. Pinto-Coelho; T. Ruão and S. Marinho (eds.), Dinâmicas comunicativas e transformações sociais. Atas das VII Jornadas Doutorais em Comunicação & Estudos Culturais (pp. 5–8). Braga: CECS.

---

[8]In addition, we may also be faced with the fact that significant science communication studies may be published outside science communication peer-reviewed publications or not even published at all due to the lack of impact or interest for practice, which the present study (especially the analysis of the proceedings book) also indicates.

**References**

Aikenhead, G. S. (2001). 'Science communication with the public: a cross-cultural event'. In: Science communication in theory and practice. Ed. by S. M. Stocklmayer, M. M. Gore and C. Bryant. Vol. 14. Contemporary trends and issues in science education. Dordrecht, The Netherlands: Springer. https://doi.org/10.1007/978-94-010-0620-0_2.

Allum, N. (2011). 'What makes some people think astrology is scientific?' *Science Communication* 33 (3), pp. 341–366. https://doi.org/10.1177/1075547010389819.

Basu, S. J. and Barton, A. C. (2010). 'A researcher-student-teacher model for democratic science pedagogy: connections to community, shared authority, and critical science agency'. *Equity & Excellence in Education* 43 (1), pp. 72–87. https://doi.org/10.1080/10665680903489379.

Bauer, M. W. (2008). 'Paradigm change for science communication: commercial science needs a critical public'. In: Communicating science in social contexts: new models, new practices. Ed. by D. Cheng, M. Claessens, T. Gascoigne, J. Metcalfe, B. Schiele and S. Shi. Dordrecht, The Netherlands: Springer, pp. 7–25. https://doi.org/10.1007/978-1-4020-8598-7_1.

— (2009). 'The evolution of public understanding of science — discourse and comparative evidence'. *Science, Technology and Society* 14 (2), pp. 221–240. https://doi.org/10.1177/097172180901400202.

Bauer, M. W., Allum, N. and Miller, S. (2007). 'What can we learn from 25 years of PUS survey research? Liberating and expanding the agenda'. *Public Understanding of Science* 16 (1), pp. 79–95. https://doi.org/10.1177/0963662506071287.

Bauer, M. W. and Jensen, P. (2011). 'The mobilization of scientists for public engagement'. *Public Understanding of Science* 20 (1), pp. 3–11. https://doi.org/10.1177/0963662510394457.

Besley, J. C. and Nisbet, M. (2013). 'How scientists view the public, the media and the political process'. *Public Understanding of Science* 22 (6), pp. 644–659. https://doi.org/10.1177/0963662511418743.

Bik, H. M. and Goldstein, M. C. (2013). 'An introduction to social media for scientists'. *PLoS Biology* 11 (4), e1001535. https://doi.org/10.1371/journal.pbio.1001535.

Brossard, D. and Schefeule, D. A. (2013). 'Science, new media and the public'. *Science* 339 (6115), pp. 40–41. https://doi.org/10.1126/science.1232329.

Bubela, T., Nisbet, M. C., Borchelt, R., Brunger, F., Critchley, C., Einsiedel, E., Geller, G., Gupta, A., Hampel, J., Hyde-Lay, R., Jandciu, E. W., Jones, S. A., Kolopack, P., Lane, S., Lougheed, T., Nerlich, B., Ogbogu, U., O'Riordan, K., Ouellette, C., Spear, M., Strauss, S., Thavaratnam, T., Willemse, L. and Caulfield, T. (2009). 'Science communication reconsidered'. *Nature Biotechnology* 27 (6), pp. 514–518. https://doi.org/10.1038/nbt0609-514.

Bucchi, M. (2017). 'Credibility, expertise and the challenges of science communication 2.0'. *Public Understanding of Science* 26 (8), pp. 890–893. https://doi.org/10.1177/0963662517733368.

Bucchi, M. and Trench, B. (2014). 'Science communication research: themes and challenges'. In: Routledge handbook of public communication of science and technology. Ed. by M. Bucchi and B. Trench. 2nd ed. London, U.K. and New York, U.S.A.: Routledge, pp. 1–13. https://doi.org/10.4324/9780203483794.

**Authors**  Sara Anjos is a Ph.D. candidate jointly at the University of Minho and at the University of Leiden. Her research interests focus on Science-Technology-Society Studies, particularly on public engagement with Astronomy. Holds a four years university degree in Astronomy, an MSc in Science Education and an MBA. She is an active member of several research groups in the Science Communication and Education intersection and was the Portuguese Language Office of Astronomy for Development — International Astronomical Union (PLOAD-IAU) coordinator until 2017. ORCID: https://orcid.org/0000-0002-8544-7471.
E-mail: saraanjos@gmail.com.

Pedro Russo is assistant professor of Astronomy & Society at Leiden Observatory and the Department of Science Communication & Society and coordinator of the Astronomy & Society group. Pedro is the president of the International Astronomical Union Commission on Communicating Astronomy with the Public. He was the global coordinator for the largest network ever in Astronomy, the International Year of Astronomy 2009.
ORCID: http://orcid.org/0000-0002-8589-7800. E-mail: russo@strw.leidenuniv.nl.

Anabela Carvalho (Ph.D., University College London) is Associate Professor at the Department of Communication Sciences of the University of Minho, Portugal. Her research focuses on various forms of environment, science and political communication with a particular emphasis on climate change. Her publications include 'Communicating climate change: discourses, mediations and perceptions' (2008), 'Citizen voices: enacting public participation in science and environment communication' (with L. Phillips and J. Doyle; 2012), 'Climate change politics: communication and public engagement' (with T.R. Peterson; 2012). She is Director of the Ph.D. programme on Communication Studies: Technology, Culture and Society. ORCID: https://orcid.org/0000-0002-7727-4187.
E-mail: carvalho@ics.uminho.pt.